\documentclass[sigconf]{acmart}

\renewcommand\footnotetextcopyrightpermission[1]{} 
\AtBeginDocument{%
  \providecommand\BibTeX{{%
    \normalfont B\kern-0.5em{\scshape i\kern-0.25em b}\kern-0.8em\TeX}}}

\usepackage{subfigure}

\setcopyright{acmcopyright}
\copyrightyear{2022}
\acmYear{2022}
\acmDOI{XXXXXXX.XXXXXXX}

\acmConference[DLP-KDD]{The 4th International Workshop on Deep Learning Practice for High-Dimensional Sparse and Imbalanced Data with KDD 2022}{August 14th-18th, 2022}{Washington DC, U.S.}

\acmPrice{15.00}
\acmISBN{XXX-X-XXXX-XXXX-X/XX/XX}

\begin{document}

\title{Entire Space Learning Framework: Unbias Conversion Rate Prediction in Full Stages of Recommender System}

\author{Shanshan Lyu, Qiwei Chen, Tao Zhuang, Junfeng Ge}
\affiliation{%
    \country{Alibaba Group, China}
}
\email{{lss271346,chenqiwei.cqw,zhuangtao.zt,beili.gjf}@alibaba-inc.com}
\renewcommand{\shortauthors}{Shanshan Lyu, et al.}


\begin{abstract}
  Recommender system is an essential part of online services, especially for e-commerce platform. Conversion Rate (CVR) prediction in RS plays a significant role in optimizing Gross Merchandise Volume (GMV) goal of e-commerce. However, CVR suffers from well-known Sample Selection Bias (SSB) and Data Sparsity (DS) problems. Although existing methods ESMM and ESM$^2$ train with all impression samples over the entire space by modeling user behavior paths, SSB and DS problems still exist. In real practice, the online inference space are samples from previous stage of RS process, rather than the impression space modeled by existing methods. Moreover, existing methods solve the DS problem mainly by building behavior paths of their own specific scene, ignoring the behaviors in various scenes of e-commerce platform. In this paper, we propose Entire Space Learning Framework: Unbias Conversion Rate Prediction in Full Stages of Recommender System, solving SSB and DS problems by reformulating GMV goal in a novel manner. Specifically, we rebuild the CVR on the entire data space with samples from previous stage of RS process, unifying training and online inference space. Moreover, we explicitly introduce purchase samples from other scenes of e-commerce platform in model learning process. Online A/B test and offline experiments show the superiority of our framework. Our framework has been deployed in rank stage of Taobao recommendation, providing recommendation service for hundreds of millions of consumers everyday.
\end{abstract}

\ccsdesc[500]{Information systems~Recommender systems}

\keywords{Recommender System, Conversion Rate Prediction}



\maketitle

\section{Introduction}

\begin{figure}
\centering
\includegraphics[width=0.4\textwidth]{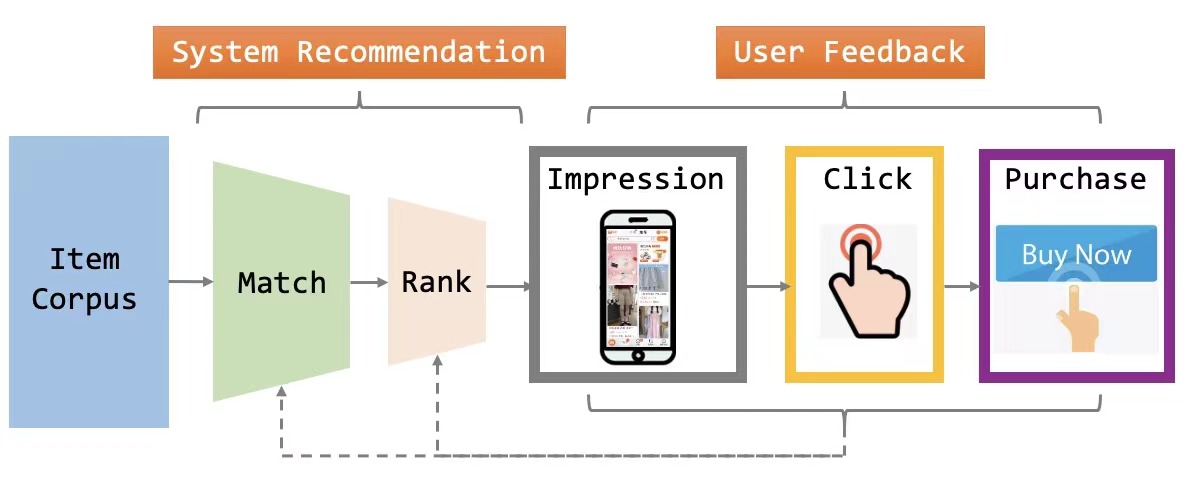} 
\vskip -10pt
\caption{Recommendation Process.}   
\label{fig:RS_process}  
\vskip -10pt
\end{figure}

Nowadays, the Internet has become an important part of People's Daily life, providing a variety of services. As the information becoming overloaded, Recommender System (RS) appears to provide personalized, timely and accurate services to users. In e-commerce platform, recommendation consists of multiple stages, i.e., match, rank and so on, as shown in Figure~\ref{fig:RS_process}. RS first recalls a large number of potential items that users may interested in according to users’ history feedback. Then, items are ranked by several metrics, such as Click-Through Rate (CTR)~\cite{2017DIN,2019DIEN}, Conversion Rate (CVR)~\cite{2018CVR,2019CVR}, etc. Finally, top items are exposed to users and users give feedback to RS, such as click, purchase. 

In this paper, we focus on CVR prediction task, which is a crucial metric in Recommender System. For example, in the e-commerce platform, maximizing GMV is one of the most important goals, 
where CVR is a significant factor. Moreover, accurately estimating CVR contribute to balance click and purchase preferences, providing better user experience.

\begin{figure}
\subfigure[Sample selection bias problem.]{
\label{fig:train_space}   
\centering
\includegraphics[width=0.2\textwidth]{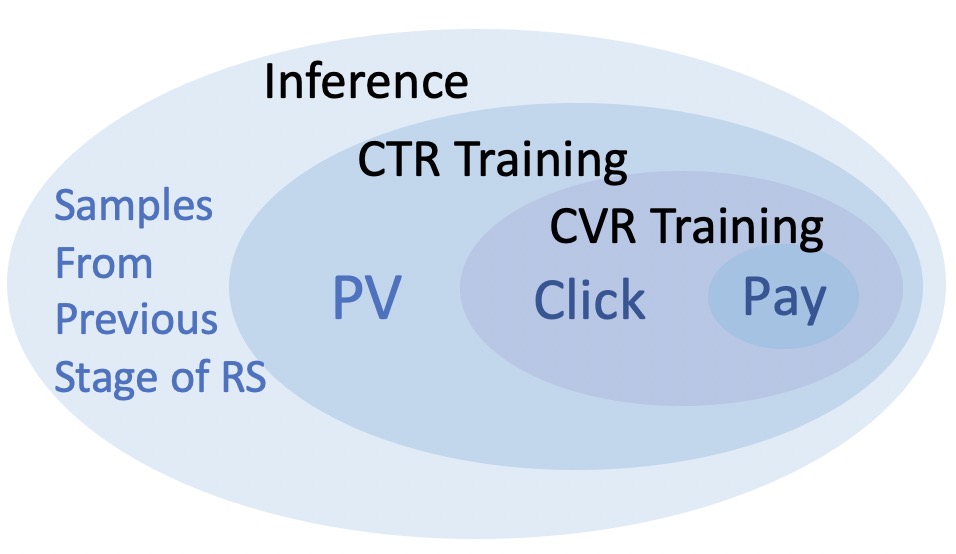} 
}
\subfigure[Various scenes in Taobao.]{
\label{fig:various_scenes}   
\centering
\includegraphics[width=0.2\textwidth]{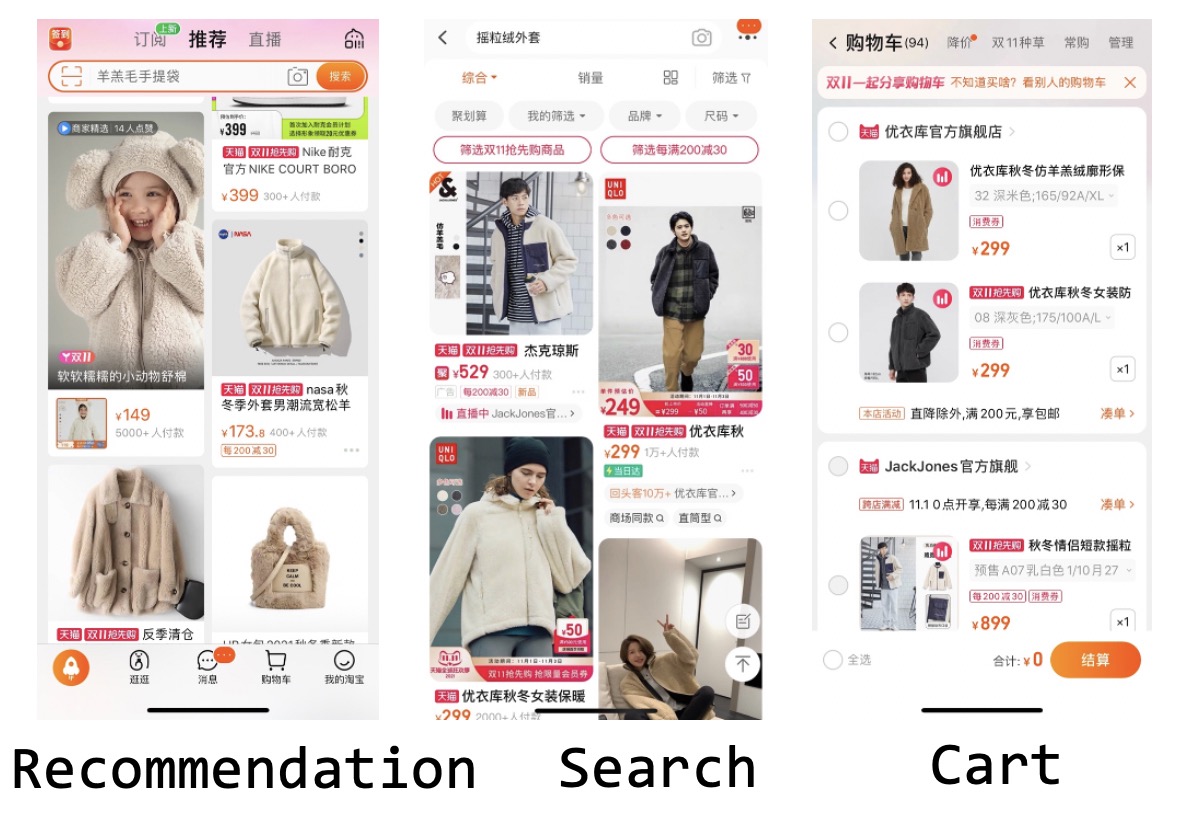} 
}
\vskip -10pt
\caption{Illustration of sample selection bias problem and various scenes in Taobao.}       
\vskip -15pt
\end{figure}

However, there exist several challenges for modeling CVR, especially two critical ones encountered in our real practice. 
1) Sample Selection Bias problem (SSB)~\cite{SSB} refers to the systematic difference between training data distribution and inference data distribution, which hurts the model performance. As shown in Figure~\ref{fig:train_space}, the inference space of rank are samples from entire space, but the training space of conventional CVR models are clicked samples, which will degrade online model performance. 
2) Data Sparsity (DS)~\cite{DSandintro} problem refers to that the number of training samples is insufficient to fit the large parameter space of models. Especially for CVR task, the training samples are a small part, which can not capture user purchase preference accurately.


Many approaches have been proposed for above challenges~\cite{2018ESMM,2020ESM,introrelated2008One,introrelated2016Bid}. For example, Entire Space Multi-Task Model (ESMM)~\cite{2018ESMM} is proposed to train CVR with all impression samples as the entire space by modeling post-view CTR and post-view clickthrough conversion rate (CTCVR) through multi-task learning framework, which alleviates SSB and DS problem.
ESM$^2$~\cite{2020ESM} employs more behaviors between click and purchase to deal with DS problem. 
However, these approaches still suffer from SSB and DS problem. 
On one hand, these methods solve SSB problem by training with all impression samples, but in real online practice the inference space of rank are samples recalled from previous match stage, as shown in Figure~\ref{fig:RS_process}. The difference between training space and inference space still exists, which will hurt model performance. 
On the other hand, these methods alleviate DS problem by building kinds of behavior paths from  click to purchase, such as cart or collect. They  ignore behaviors in various scenes of e-commerce platform. 
As shown in Figure~\ref{fig:various_scenes}, in Taobao\footnote{https://www.taobao.com/}, one of the largest e-commerce platforms in China, click and purchase behaviors happen in a variety of scenes, i.e. search, recommendation, cart and so on. Users’ feedback in all scenes of e-commerce platform can help capture users’ preference more accurately. Thus, how to utilize users’ behaviors of all scenes in e-commerce platform is a very important issue.

In this paper, we propose Entire Space Learning Framework(ESLM): Unbias Conversion Rate Prediction in Full Stages of Recommender System to deal with SSB and DS problems. We propose a novel formulation of CVR on the entire space by reformulating the GMV goal from the overall view of recommender systems. 
Specifically, we train CVR model by formulating the probability on the entire data space from previous stage of RS to purchase behavior in our own scene. Therefore, the difference between training space and inference space is disappeared. 
Moreover, we model users’ purchase feedback in our own scene as well as other scenes of Taobao explicitly, alleviating the DS problem. Specifically, we decompose CVR into two parts: 1) probabilty from entire data space to purchase behavior in all scenes and 2) probabilty from purchase samples in all scenes to purchase samples our own scene.
Offline experiments and online A/B test show the superiority of our framework comparing to existing methods. Our framework has been deployed in rank stage of Taobao recommendation, which provides recommending service for hundreds of millions of consumers everyday.

\section{Framework}


\subsection{Formulation}
Recommender system of e-commerce consists of several stages, mainly including match and rank. In the match stage, a small number of items are quickly retrieved from the massive candidates according to users’ characteristics. Then items from the match stage are as input to the rank stage. One of the most important optimization objectives of e-commerce platform is the GMV which can be decomposed into 
\begin{equation}
GMV = Traffic * CTR * CVR * price.
\end{equation}

Traditional CVR models probability $p_{ClickToPay} = p(Pay=1|Click=1)$ for a sample. ESMM propose to train CVR with all impression samples over the entire space by modeling CTCVR as $p_{PvToPay} = p(Click=1|Pv=1)*p(Pay=1|Click=1, Pv=1)$.
However, there are two problems with this modeling. 1) Sample selection bias. The data space of training and inference is different, where the training space are impression samples or clicked samples but the inference space are samples from previous stage of RS. 2) Data sparsity problem. Traditional model only trains on samples in our own scene, ignoring the rich behaviors in all scenarios of Taobao, which causes data sparsity problem. Therefore, how to keep the training and inference space consistent, and how to utilize samples from all scenes of Taobao in all stages of RS are very important issues.

Therefore, we propose an entire space learning framework 
reconstructing the probability space as a supplement of the existing GMV function to alleviate above problems. 
First, we rebuild the probability space. In order to make the training space consistent with the online inference space, we make full use of the recommender system structure, directly constructing the probability of samples from previous stage to purchase samples of our scene, defined as 
\begin{equation}
    p_{PSToPay_{g}} = p(Pay_{g}=1|PS=1).
\end{equation}
where $\{Pay_{g}=1\}$ is the purchase samples in our own scene and $\{PS=1\}$ is the samples from previous stage of RS (as shown in Figure~\ref{fig:RS_process}). Furthermore, in order to alleviate the data sparsity problem, we employ purchase samples in all scenes of Taobao at each stage of RS, i.e., match and rank. We further decompose the probability into the following form.
\begin{equation}
\begin{aligned}
    p_{PSToPay_{g}} = & p(Pay_{g}=1|PS=1) \\
                = & p(Pay_{a}=1|PS=1) *  p(Pay_{g}=1|Pay_{a}=1, PS=1).
\end{aligned}
\end{equation}
where $\{Pay_{a}=1\}$ is the joint set of samples from previous stage and purchase samples in all scenes of Taobao, and probability ${p(Pay_{g}=1|Pay_{a}=1, PS=1)}$ is the probability of purchase samples in all scenes to purchase samples in our own scene. 

This modeling manner of CVR has three advantages. 1) The candidate data space of training and inference are consistent. The offline training and online inference data space are samples returned from previous stage in RS. Hence there is no inconsistency between online prediction and offline training, alleviating sample selection bias problem. 2) We explicitly introduce purchase samples of all scenes in Taobao in the model learning process, which alleviates data sparsity problem. However, merely introducing purchase samples of other scenes to our model may expose more items not fit our scene. 3) We modeling the probability of purchase sample in all scenes to our own scene, as a result of which the final modeling target is the positive sample of our own scene. This modeling manner is more in line with the mind of our own scene, reducing exposure of samples which is not suitable for our own scene.

Finally, it is used as a supplement of traditional optimization goal GMV, formalized as:
\begin{equation}
    GMV = Traffic * (p_{PvToPay} + \alpha * p_{PSToPay}) * price.
\end{equation}

\subsection{Model}

\begin{figure}
\centering
\includegraphics[width=0.42\textwidth]{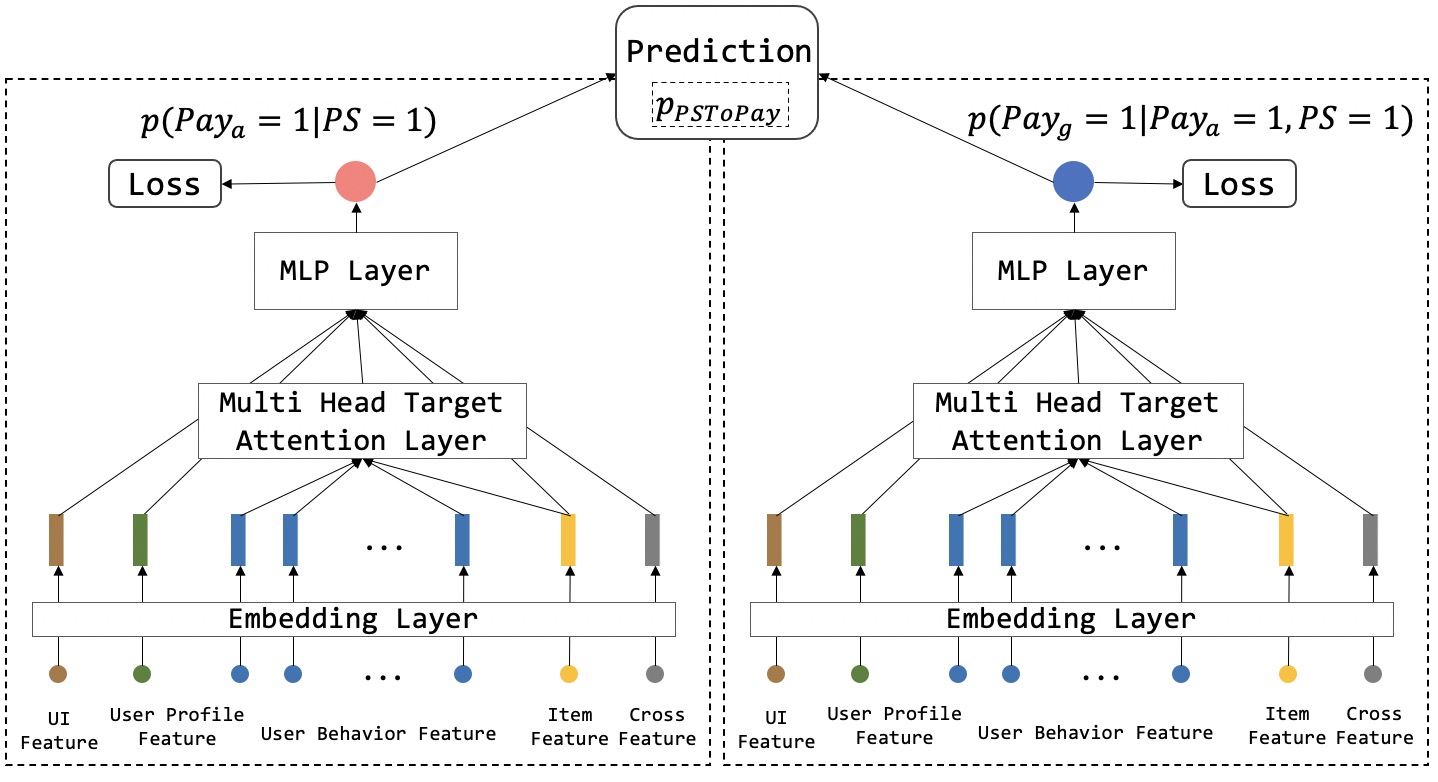} 
\vskip -10pt
\caption{The network of our model.}   
\label{fig:network}     
\vskip -15pt
\end{figure}

In this section, we introduce our model network in detail ( Figure~\ref{fig:network}).
The model consists of two parts. The first part $p(Pay_{a}=1|PS=1)$ models the probability of samples from previous stage to purchase samples in all scenes. The second part $p(Pay_{g}=1|Pay_{a}=1, PS=1)$ models the probability from purchase samples in all scenes to purchase samples in our own scene.

\textbf{Embedding Layer} maps a large number of sparse features into low-dimensional and dense vector space. The input features include user profile feature, user behavior sequence, item attribute feature, and context feature.

\textbf{Multi Head Target Attention Layer} captures the similarity between the target item and user's historical behavior sequence. The standard TA is based on the structure of multi-head attention~\cite{2017Attention} and has been widely applied to sequential user data modeling in CTR models~\cite{2017CTRTA,2020CTRTA,2020CAN}. Different from NLP models, CTR models treat the target item as Query and the user behavior sequence as Key and Value. The formulation of single-head attention and the representation of user sequence is formalized as
{\setlength\abovedisplayskip{0pt}
\begin{equation}
\begin{aligned}
    & Attention_i(Q, K, V) = softmax(\frac{QK^{T}}{\sqrt{d_k}}), \\
    & S = Concat(Attention_1, Attention_2, \dots, Attention_h)W^S,
\end{aligned}
\end{equation}
\setlength\belowdisplayskip{0pt}} where $Q=E_sW_i^Q, K=E_sW_i^K, V=E_sW_i^V$ with $Es$ denoting the embedding matrices of items, and $W_i^Q, W_i^K, W_i^V$ denoting the linear projection matrices of $i-th$ head, $Concat(\cdot)$ denotes the concatenation operation and $W^S$ denotes the linear projection matrix with $h$ denoting the number of heads.

\textbf{MLP layers and Loss function}. By concatenating the representation of all features, we use three fully connected layers to further learn the interactions among dense features, which is widely used in industrial RS.

In order to predict whether a user will buy the target item in all scenes of Taobao, we model it as a binary classification problem by employing $Sigmoid$ function as the output unit. Finally, the cross-entropy loss function is defined as 
{\setlength\abovedisplayskip{0pt}
\begin{equation}
\begin{aligned}
    &Loss_{PSToPay_a} = - \frac{1}{N} \sum_{(x,y) \in D}^N (ylogp(x)) + (1-y)log(1-p(x))), \\ 
\end{aligned}
\end{equation}
\setlength\belowdisplayskip{0pt}} where $p(x)=p(Pay_a=1|PS=1, x)$ represents the predicted probability of sample $x$ bought by the user in all scenes of Taobao, $D$ is the dataset of all the samples from previous stage of RS, and $y \in {0, 1}$ is the label indicating whether the target item is bought by the user in all scenes of Taobao.

Similarly, in order to model whether a user will buy the target item in our scene or other scenes, the loss function is defined as:
{\setlength\abovedisplayskip{0pt}
\begin{equation}
\begin{aligned}
    Loss_{Pay_aToPay_g} &= - \frac{1}{N} \sum_{(x,y) \in D_{p}}^N (ylogp(x)) + (1-y)log(1-p(x))), 
\end{aligned}
\end{equation}
\setlength\belowdisplayskip{0pt}} where $p(x)=p(Pay_a=1|PS=1, x)$ represents the predicted probability of sample bought in our scene or other scenes, $D_{p}$ is the joint dataset of samples from previous stage of RS and purchase samples in all scenes, and $y \in {0, 1}$ is the label indicating whether the target item is bought in our scene.

\section{Experiment}

\subsection{Experimental Setups}
\textbf{Dataset.} The dataset is constructed from the log of Taobao App\footnote{https://www.taobao.com/}. We construct an offline dataset based on users’ behaviors in fifteen days. We use the first fourteen days as training data, and the last day as test data. The statistics of the dataset is shown in Table~\ref{tab:dataset}, which is extremely large and sparse. 
Traditional training dataset is from impression logs with statistics shown in Table~\ref{tab:dataset}, which can not satisfy our novel paradigm. Thus, we construct a new dataset. First, we log all samples from previous stage of ranking $\{PS=1\}$ and log purchase samples in all scenes of Taobao $\{Pay=1\}$. Then, we join $\{PS=1\}$ and $\{Pay=1\}$ to utilize the purchase samples $\{Pay_a=1\}$. We use negative sampling for negative samples. Dataset $\{Pay_a=1\}$ can be devided into positive samples in our scenario $\{Pay_g=1\}$ and others. $\{Pay_a=1\}$ brings mutiple times of positive samples to our model which is effective for data sparsity problem.

\textbf{Evaluation metric.} For online A/B test, we use CVR to evaluate all models.
For offline evaluation, we use Area Under Curve (AUC) metric. We evaluate AUC with several kinds of lables on both datasets. We evaluate AUC(PvToPay$_g$) on $\{Pv=1\}$ impression dataset with $label=1$ if the item is purchased in our scene, AUC(PSToPay$_g$) on $\{PS=1\}$ dataset with $label=1$ if the item is purchased in our scene, and AUC(PSToPay$_a$) on $\{PS=1\}$ dataset with $label=1$ if the item is purchased in all scenes of Taobao. 
Traditional offline evaluating Area Under Curve (AUC) is hard since our framework will change the impression items. Thus, we need to analyze by combining online and offline evaluation results.

\textbf{Baselines.} To evaluate the effectiveness of our model, we compare it with ESMM\cite{2018ESMM}, ESM$^2$\cite{2020ESM}, and our baseline method which trains cvr and ctr separately on our own scene data. 
In order to validate the effectiveness of unifying the train and inference data space and explicitly employing purchase samples in other scenes, we introduce variants of our model which are as follows.
1) Pv2Pay$_g$ is trained on the impression dataset, which modeled the probability $p(Pay_g=1|Pv=1)$. 
2) PS2Pay$_g$ is trained on samples from previous stage, modeling the probability $p(Pay_g=1|PS=1)$, which does not introduce samples of other scenes compared to our framework.

\textbf{Settings.} Our model is implemented with Python 2.7 and Tensorflow 1.4, and the "Adagrad" is chosen as the optimizer with learning rate 0.01. The hyparameter alpha is finetuned by gridsearch according to offline auc and online A/B test and we finally use 0.1.

\subsection{Experimental Results}

In this section, we analyze experimental results aiming to answer three questions. RQ1: Does our framework perform better than existing methods in CVR prediction? RQ2: Does our method alleviate sample selection bias by modeling based on $\{PS=1\}$ dataset? RQ3: Is it effective to alleviate DS problem by introducing users' behaviors in other scenarios of Taobao explicitly?

\begin{table}
\centering
\caption{Dataset statistics.}
\label{tab:dataset}
\vskip -8pt
\begin{tabular}{c|c c c}
\hline
Dataset &  Users &  Items & Samples\\
\hline
Taobao & 463,734,383 & 136,025,054 & 90,849,473,690\\
\hline
\end{tabular}
\vskip -15pt
\end{table}

To answer RQ1, we compared our model with ESMM, ESM$^2$. From online results (Table~\ref{tab:auc1}), it is observed that our model performs better than existing state-of-the-art methods,  proving the superiority of our framework in solving SSB and DS problem. From Table~\ref{tab:auc1}, the auc of our model on existing impression dataset is a little lower but the online CVR increase is significantly higher. This is because that our model changes the online data distribution and is not consistent with existing dataset space.

To answer RQ2, we compared variants of our model PvToPay$_g$ and PSToPay$_g$. From Tabel~\ref{tab:auc2}, it is observed that the PSToPay$_g$ has higher AUC and better online results than PvToPay$_g$, proving the effectiveness in alleviating SSB problem.

To answer RQ3, we compared whether to introduce samples of other scenarios in our framework. From Tabel~\ref{tab:auc1} and Tabel~\ref{tab:auc2}, it is observed that the experimental results of our framework are significantly better than those of the models without introducing global samples. It demonstrates that introducing global domain knowledge by the way of our framework alleviates data sparsity problem and improves the generalization performance of the model, which results in better performance in ranking.


\begin{table}
\centering
\caption{Online CVR gains and offline AUC(PvToPay$_g$) on $\{Pv=1\}$ impression dataset of state-of-the-art methods.}
\label{tab:auc1}
\vskip -8pt
\begin{tabular}{c|c|c}
\hline
Method & AUC(PvToPay$_g$) & CVR Gain\\
\hline
Baseline & 0.8623 & - \\
ESMM & 0.8634 & +1.3\% \\
ESM$^2$ & 0.8643 & +2.2\% \\
Our & 0.8548 & +7.3\%\\
\hline
\end{tabular}
\vskip -10pt
\end{table}

\section{Related Work}

\textbf{Conversion Rate Prediction.} There are few literatures directly proposed for CVR task~\cite{2018ESMM,2020ESM,related16,related2016Large,related2019Multi}.
ESMM~\cite{2018ESMM} and ESM$^2$~\cite{2020ESM} are proposed for SSB and DS problem in CVR task. ESMM adds the CTR task and CTCVR task as an auxiliary to the main CVR task. ESM$^2$ propose postclick behavior decomposition and formulate the final CVR as well as some auxiliary tasks together. As a result, these two methods address the SSB and DS issue simultaneously by building user behavior paths. However, these methods still suffers from SSB and DS problems. In the real recommender system, the online inference space are samples from previous stage of RS, but not the impression space which is used in these two mehods. For DS problem, these two methods model behaviors of their own scene, ignoring  behaviors of other scenes. Our framework alleviate the SSB and DS problem by reformulating the purchase probability of the specific scene. Our model construct the probability on samples from previous stage of RS, unifying the training space and inference space. For DS problem, we employ purchase samples in all scenes of Taobao by decomposing the probability. Such that our model can leverage all samples over the entire space and purchase labels from all scenes of Taobao, consequently addressing the SSB and DS issue simultaneously. 

\textbf{Multi-task Learning.} Although multitask learning framework (MMOE~\cite{2018MMOE}, PLE~\cite{2020PLE}) can model the same task of different scenarios in a multitask learning minor, as two towers CVR in scenario $A$ and CVR in scenario $B$. The final output is still the CVR in scenario $A$, which utilize the label sample in an implicit and impact the embedding and NN weights of the main task. We utilize the label sample in an explicit minor, which is more suitable for online deployment.

\begin{table}
\centering
\caption{Online CVR gains and offline AUC(PSToPay$_g$), AUC(PSToPay$_a$) on $\{PS=1\}$ dataset of variants of our model.}
\label{tab:auc2}
\vskip -8pt
\begin{tabular}{c|c|c|c}
\hline
Method & AUC(PSToPay$_g$) & AUC(PSToPay$_a$) & CVR Gain\\
\hline
Baseline & 0.8623 & 0.9325 & - \\
Pv2Pay$_g$ & 0.8574 & 0.9286 & - \\
PS2Pay$_g$ & 0.8233 & 0.9514 & +0.8\% \\
Our & 0.8548 & 0.9562 & +7.3\%\\
\hline
\end{tabular}
\vskip -5pt
\end{table}

\section{Conclusion}
In this paper, we propose a novel framework of modeling GMV in Taobao. The model unifies the train and inference space and introduces label samples in all scenes of Taobao, alleviating the SSB and DS problem.
Online and Offline experimental results show the superiority of our framework comparing to existing methods. Our framework has been deployed in ranking stage for Taobao homepage recommendation, providing services for hundreds of millions of users in China.
For future work, the framework can extend to other tasks such as CTR.


\bibliographystyle{ACM-Reference-Format}
\bibliography{my_ref}

\end{document}